\documentclass[11pt]{article}
\usepackage{authblk}

\usepackage{mathtools,amsmath,graphicx}
\usepackage[round,numbers, sort&compress]{natbib}

\DeclareMathOperator*{\argmin}{\arg\!\min}
\makeatletter
  \def\tagform@#1{\maketag@@@{[#1]\@@italiccorr}}
\makeatother
\usepackage{setspace}
\usepackage[margin=1in]{geometry}
\usepackage{amssymb}
\usepackage{epstopdf}
\usepackage[hang]{subfigure}
\usepackage{color}
\usepackage{amsthm}
\usepackage{algorithm}
\usepackage[percent]{overpic}
\usepackage{graphicx}
\usepackage{algorithmic}
\usepackage{url}
\title{\uppercase{ \bf \large \center  Improved Reconstruction for high-resolution Multi-shot Diffusion Weighted Imaging}}
\author[1]{\it \small Merry Mani}
\author[2]{\it \small Hemant Kumar Aggarwal}
\author[1]{\it \small Vincent Magnotta}
\author[2]{\it \small Mathews Jacob}
\affil[1]{\it \small Department of Radiology, University of Iowa, Iowa City, Iowa}
\affil[2]{\it \small Department of Electrical and Computer Engineering, University of Iowa, Iowa City, Iowa}

\begin{document}
\maketitle

\vspace{30mm}
\noindent  Correspondence to :\\
Merry Mani\\
L309 Papajohn Biomedical Discovery Building,\\
169 Newton Road\\
Iowa City, Iowa, 52242\\

\noindent  email: merry-mani@uiowa.edu \\
phone number: (319) 335-9569.\\
\\

Word count : about 5000\\
figures+ tables count : 10\\
\\

\newpage
\noindent { \bf Abstract}\\

\noindent {\bf Purpose:} To introduce a fast and improved direct reconstruction method for multi-shot diffusion weighted (msDW) scans for high-resolution studies.\\

\vspace{-1.5em}\noindent {\bf Methods:} Multi-shot EPI methods can enable higher spatial resolution for diffusion MRI studies. Traditionally, such acquisitions required specialized reconstructions involving phase compensation to correct for inter-shot motion artifacts. The recently proposed MUSSELS reconstruction belongs to a new class of parallel imaging-based methods that recovers artifact-free DWIs from msDW data without needing phase compensation. 
However, computational demands of the MUSSELS reconstruction scales as the matrix size and the number of shots increases, which hinders its practical utility for high-resolution applications.  In this work, we propose a computationally efficient formulation using iterative reweighted least squares (IRLS) method. The new formulation is not only fast but it enables to accommodate additional priors such as conjugate symmetry property of the k-space data to improve the reconstruction. Using whole-brain in-vivo data, we show the utility of the new formulation for routine high-resolution studies with minimal computational burden.\\

\vspace{-1.5em}\noindent {\bf Results:}  
 The IRLS formulation provides about six times faster reconstruction for matrix sizes 192$\times$192 and 256$\times$256, compared to the original implementations. The reconstruction quality is improved by the addition of conjugate symmetry priors that reduce blurring and preserves the high-resolution details from partial Fourier acquisitions. \\

\vspace{-1.5em}\noindent {\bf Conclusion: } The proposed method is shown to be computationally efficient to enable routine high-resolution studies. The computational complexity matches the traditional msDWI reconstruction methods and provides improved reconstruction results.

\newpage

\noindent {\bf \Large Introduction}\\ 

\vspace{-1.5em}\noindent The spatial resolution limits of diffusion weighted images (DWIs) has been traditionally set by the capabilities of single-shot echo planar imaging (ssEPI) techniques on a given set of MRI hardware. On standard clinical gradients (33mT/m gradient strength, 120T/m/s slew rate), ssEPI readouts contribute to a minimum TE of around 75-120 ms for a 128 x 128 imaging matrix for typical b-values of 1000-3000 $s/mm^2$. Pushing the spatial resolution of DWIs beyond the 128 x 128 matrix size results in (i) increased geometric distortions along the phase-encoding direction, (ii) reduced signal-to-noise ratio (SNR) due to the long read out time involved in sampling the center-of-k-space and (iii) increased blurring in the images due to the T$_2$* signal decay that accompanies the long readout duration \cite{Wu2017}. Thus, despite not being the ideal resolution to study the {\it{micro-level}} tissue structural properties, the characteristic resolution of standard DWI studies has largely remained at approximately 2mm isotropic. On the other hand, during a practical diffusion encode time ($\delta$), the diffusion measurements are sensitive to length scales of the order of $\sim$10$\mu$m \cite{Novikov2018b}. However, the robust detection of subtle micro-structural changes of order $\sim$10$\mu$m length scale from a 2$mm$ voxel resolution will result in significant partial volume artifacts.
Thus, there is a strong biological motivation to push the spatial resolution of DWIs to sub-millimeter scales \cite{Jones2018}. Multi-shot echo-planar imaging (msEPI) readouts can enable higher spatial resolutions for DWIs on standard clinical MRI hardware. Combined with synergistic improvements offered by high field strengths, high performance gradients, and high slew rate systems, embracing msEPI methods for DWIs can set the stage to push the spatial resolution of DWIs to sub-millimeter voxel sizes. \\

\vspace{-1.5em}\noindent A major concern while employing msEPI methods for diffusion studies is the fact that the reconstruction of multi-shot diffusion weighted (msDW) data is not amenable to the traditional IFFT-based reconstructions. This is because the k-space of a {\it given DWI} is sampled over multiple TRs using a msEPI scheme which contributes to a unique phase for the data corresponding to the different readouts \cite{Anderson1994,Eichner2015a}. Thus, the data from the k-space segments acquired across multiple TRs need to be compensated for their inconsistent phases before they can be combined. Traditionally, this phase compensated reconstruction involved a multi-stage reconstruction of the individual shot images to calibrate their phases and finally combining the images using phase compensation methods \cite{Wu2017,Butts1996,Chen2013a}. Recently, several new strategies that can combine the k-space data from multiple shots without needing explicit phase compensation have been proposed \cite{Mani2017a,Mani2016d,Hu2018a}.  Such reconstructions can enable direct recovery of the DWIs from the k-space data and thus can potentially enable online reconstruction of the DWIs on the scanner. One outstanding challenge with the above iterative reconstructions compared to the IFFT-based reconstructions is the long reconstruction times involved. \\

\vspace{-1.5em}\noindent In this work, we focus on the MUSSELS (Multi-shot sensitivity-encoded recovery using structured low-rank matrix completion) reconstruction \cite{Mani2017a} and identify some of its computational bottlenecks. We propose a computationally efficient formulation to solve the matrix recovery problem embedded in the above msDW data recovery scheme. The proposed formulation offers the following advantages: (i) the computation time is reduced by several folds for the recovery of high-resolution data, (ii) the improved efficiency allows to accommodate additional constraints to achieve more accurate image recovery without trading off computational time. The proposed formulation is shown to provide high-quality DWI reconstruction with highly consistent image recovery from the multiples shots.  We show reconstruction results of high-resolution DWI data that show improved recovery of anatomical details without incurring high computational burden. We also show the feasibility of whole-brain high-resolution studies using the proposed reconstruction and highlight the benefits of the higher resolution in comparison to standard 2mm isotropic dMRI data.\\

 \noindent {\bf \Large Theory }\\

\vspace{-1.5em} \noindent { \large \it Direct reconstruction of msDWI using MUSSELS}\\ 

\vspace{-1.5em} \noindent A pictorial representation of the MUSSELS reconstruction proposed in \cite{Mani2017a} is given in figure 1. This method is notably different from the traditional msDW reconstruction methods  (e.g. navigator-based methods \cite{Holdsworth2008,Porter2009}, self-navigated methods such as \cite{Liu2004, Guo2016} including MUSE \cite{Chen2013a,Chu2015}) that involves phase compensation.  
Specifically, the MUSSELS reconstruction does not involve phase compensation and aims to recover the missing k-space data in each of the shots jointly as shown in figure 1. The recovery is performed in a regularized parallel imaging setting, where the regularization is provided by a low-rank prior. The low-rank prior is derived from the several annihilation relationships that exist between the msDW data. The annihilation relations emerge from the fact that the DWIs corresponding to each shot of the multi-shot acquisition, are the phase modulated versions of the same underlying DW magnitude data (see figure 1). This is because the diffusion gradients encode the microscopic motion (diffusing motion, pulsatory, respiratory etc.) in the phase of the complex acquired signal \cite{OHalloran2012}. This {\it {background phase}} \cite{Eichner2015a} varies for each TR due to the inter-TR microscopic motion, whereas the magnitude of the signal attenuation remains the same for a given diffusion encoding gradient direction. Thus annihilation relations of the form \begin{equation}
	\label{first_eq}
	  m_{i}({\bf x})\phi_{j}({\bf x})-m_{j}({\bf x})~\phi_{i}({\bf x}) = 0; \forall \bf x.
	 \end{equation} 
 exist between every pair of complex shots images $i,j$ of the multi-shot acquisition of a given diffusion weighting. The various notations are defined and illustrated in figure 1.\\ 
 
 \vspace{-1.5em} \noindent The key idea of the MUSSELS reconstruction is that the multiple annihilation relations present in the multi-shot data are exploited to form a low-rank system in the {\it{frequency domain}} which enables direct reconstruction of the multi-shot k-space data. Specifically, the rank deficiency of a Hankel-structured matrix formed from the k-space data of the multi-shot acquisition provides a constraint for the multi-shot data recovery. Thus, the MUSSELS reconstruction aims to recover the missing k-space data in each k-space shot using matrix completion algorithms that enforce low rank on the Hankel-structured matrix. This allows recovery of the underlying diffusion weighted magnitude data to be estimated as the sum-of-squares (SOS) of the images reconstructed from individual shots. \\
 
 \vspace{-1.5em} \noindent  { \large \it Computational complexity of MUSSELS}\\ 

 \vspace{-1.5em} \noindent In this section, we identify some of the major computational bottlenecks of the MUSSELS reconstruction.  The low-rank prior exploited in the MUSSELS reconstruction is derived from the annihilation relations in the image domain given in Eq. \ref{first_eq}. These relations hold in the frequency domain via convolutions as: 
 \begin{equation}
	\label{sec_eq}
	   \widehat{m_{i}}[{\bf k}]*\widehat{\phi_{j}}[{\bf k}]-\widehat{m_{j}}[{\bf k}]*\widehat{\phi_{i}}[{\bf k}] = 0; \forall \bf k.
	 \end{equation} 
	 Since convolutions can be implemented mathematically as matrix multiplication using convolution matrices, the above equation can be written as a matrix multiplication of the form:
	 \begin{equation}
	\label{third_eq}
	{\mathbf{H}}(\widehat{{m}_{i}})\cdot \widehat{\boldsymbol\phi}_{j}-{\mathbf {H}}(\widehat{{m}_{j}})\cdot \widehat{\boldsymbol\phi}_{i} = \bf 0,
	\end{equation}
	 where ${\mathbf{H}}(\widehat{{m}_{i}})$ is the Hankel-structured convolution matrix. The presence of multiple annihilation relations for every pairing of $i,j \in \{1,2,\hdots,N_s\}$, $N_s$ being the number of shots, establishes the low-rank property of the block-Hankel matrix ${\bf{\bf{H_1}}}({\bf{\widehat{{m}}}})$
	 \begin{equation}
	\label{equation 3}
	\underbrace{\begin{bmatrix} {\bf{H}}(\widehat{m_1}) & {\bf{H}}(\widehat{m_2}) & ... & {\bf{H}}(\widehat{m_{N_s}}) \end{bmatrix}}_{{{\bf{H}}}_1(\bf\widehat{m})}\underbrace{\left[ \begin{array}{c} \widehat{\boldsymbol\phi_2} \\ -\widehat{\boldsymbol\phi_1}\\ 0 \\ 0\\  \vdots \\ 0 \end{array} \begin{array}{c} 0\\
 \widehat{\boldsymbol\phi_3} \\ -\widehat{\boldsymbol\phi_2}\\ 0 \\ \vdots \\ 0 \\ \end{array}\begin{array}{c} \cdots \end{array}
\begin{array}{c} 0\\0 \\ \vdots \\ 0 \\ \widehat {\boldsymbol\phi_{N_s}}\\
-\widehat {\boldsymbol\phi_{N_s-1}}  \end{array}\begin{array}{c}  \widehat{\boldsymbol\phi_3} \\0\\-\widehat{\boldsymbol\phi_1} \\ 0\\  \vdots \\ 0 \end{array}  \begin{array}{c} \cdots \end{array} \right]}_{\bf{\hat{\Phi}}}= \begin{bmatrix}0 & 0 & ...& 0 & 0 ... \end{bmatrix}, 		\end{equation}
owing to the existence of its non-trivial null-space. The MUSSELS reconstruction for the recovery of the multi-shot k-space data makes use of the above low-rank prior on ${\bf{\bf{H_1}}}({\bf{\widehat{{m}}}})$ by solving the following rank minimization problem: 
 \begin{equation}
\label{use Hankel}
\widehat{\tilde {\bf m}}=  \argmin_{{\bf{\widehat{m}}}} \underbrace{||{\mathcal A}\left({\bf{\widehat{m}}}\right)-{\bf{\widehat{y}}}||^2_{\ell_2}}_{\mbox{data consistency}} + \lambda \underbrace{||{\bf{\bf{H_1}}}({\bf{\widehat{{m}}}})||_*}_{\mbox{low-rank penalty}} .
\end{equation}
 Here, the data consistency term relies on a SENSE-based formulation \cite{Pruessmann1999} to ensure consistency to the measured k-space data $\bf{\widehat y}$. The forward operator ${\mathcal A}$ is given by $\bf{{\cal{M}}}\circ\bf{{\cal{F}}}\circ\bf{{\cal{S}}}\circ\bf{{\cal{F}}^{-1}}$ where $\bf{{\cal{F}}}$ and $\bf{{\cal{F}}^{-1}}$ represent the Fourier transform and the inverse Fourier transform operations respectively, $\bf{{\cal{S}}}$ represents multiplication by coil sensitivities, and $\bf{{\cal{M}}}$ represents multiplication by the k-space sampling mask corresponding to each shot. The second term enforces low-rank on the block-Hankel matrix by minimizing the nuclear norm. \\
 
 \vspace{-1.5em} \noindent The structure of ${\bf H_1}({\bf \hat m})$ is as defined in Eq. \ref{equation 3} and is a concatenation of the convolutional Hankel matrices corresponding to the k-space data of each shot. Using structured low-rank matrix completion, the recovery problem in Eq. \ref{use Hankel} recovers, ${\bf \hat m}$,  the matrix of k-space data from all shots, jointly. The above cost function is minimized in an iterative fashion using alternating minimization schemes.
 The augmented Lagrangian method \cite{Bertsekas1976} is a standard choice to solve such cost functions. Here, the cost function in Eq. \ref{use Hankel} is split into two sub-problems using auxiliary variable $\bf D$ as a surrogate for ${\bf H_1}({\bf \hat m})$. The first sub-problem updates $\bf \hat m$ by minimizing the quadratic cost function:  
 \begin{equation}
\label{AL2}
C_1({\bf \hat m})= {||{\bf{\cal{A}}({\bf \hat m})}-{\bf \hat y}}||^2_{\ell_2} +\frac{\lambda\beta}{2}||{\bf D}-{\bf{H_1}}({\bf \hat m})||^2_{\ell 2} +\frac{\lambda}{2}\hat{\gamma}^T({\bf D}-{\bf H_1}({\bf \hat m})), \end{equation} using conjugate gradients (CG), while the second cost function: 
  \begin{equation}
\label{AL2}
 C_2({\bf D}) = ||{\bf D}-{\bf H_1}({\bf \hat m})||^2_{\ell 2} + \frac{2}{\beta}||{ \bf D}||_*   \\
\end{equation}updates $ \bf D$ via singular value shrinkage \cite{Fornasier2011}. Here $\beta$ and $\gamma$ are respectively the penalty parameter and the Lagrange multipliers \cite{Ramani2011}.   \\

 \vspace{-1.5em} \noindent {\it Singular Value Shrinkage: }One major computational bottleneck in the above MUSSELS implementation stems from the singular value shrinkage step involving the singular value decomposition (SVD) of the block-Hankel matrix $\bf{\bf{H_1}}({\bf{\widehat{{m}}}})$. If the size of the DWI is given by $N_1 \times N_2$, then the size of the block-Hankel matrix is given by $ m \times n$ where $ m=(N_1-r+1)\times (N_2-r+1)$ and $n=r\times r\times N_s$ where $r$ is the size of the filter. Since $m >> n$ (by $\sim$ 2-3 orders of magnitude), the above minimization incurs a computational complexity of $O(mn^2)$ for every iteration of $C2$. \\

 \vspace{-1.5em} \noindent  {\it Concatenated Hankel matrix solves a multi-channel convolution: }A second bottleneck is associated with the size of the {\it lifted matrix $\bf{\bf{H_1}}({\bf{\widehat{{m}}}})$} itself which is much larger in size compared to the original data ${\bf{\widehat{m}}}$. Computing and storing the above matrix in addition to computing the SVD during every iteration makes the above algorithm computationally demanding. Complexity grows with matrix sizes, which makes it difficult to add more priors further increasing the matrix sizes. A straightforward solution to overcome the computational complexity is to work in the domain of the original data as opposed to the lifted matrix. This approach is considered in \cite{Ongie2017}, where the matrix lifting is eliminated using a half circulant approximation of the multi-level block Toeplitz matrix, which reduces the matrix sizes comparable to that of the original data. However, the above half-circulant approximation is not valid for the current setting since $\bf{\bf{H_1}}({\bf{\widehat{{m}}}})$ is not truly "block-Hankel", but rather a concatenation of Hankel matrices. Additionally,  due to the concatenation of the Hankel matrices, the current work presents a more generalized multi-channel convolution as opposed to the single-channel convolutional setting considered in \cite{Ongie2017}. Implementing multi-channel convolutions by extending the half circulant approximation of individual Hankel matrices to the current setting is computationally more demanding than working with lifted matrices. Hence, we refrain from the discrete Fourier Transform (DFT) based implementations in \cite{Ongie2017} and seek a more computationally efficient method suited to work with lifted matrices.\\

 \vspace{-1.0em} \noindent { \it \large Iterative Reweighted Least Squares formulation of MUSSELS}\\ 

 \vspace{-1.5em} \noindent In this section, we adopt a new formulation that can significantly reduce the computational burden of the MUSSELS reconstruction. The iterative reweighted least squares (IRLS) method has been applied to solve nuclear norm optimization problems by several authors where the original problem of minimizing the Schatten p norm is replaced by a re-weighted Frobenius norm \cite{Chartrand2008a,Fornasier2011,Mohan2010}. The weights for the modified problem are computed from the previous iteration and is re-weighted every iteration. Thus, the nuclear norm minimization of the form 
 \begin{equation}
  \min_{{\bf{\widehat{m}}}} ||{\mathcal A}\left({\bf{\widehat{m}}}\right)-{\bf{\widehat{y}}}||^2_{\ell_2} + \lambda ||{\bf{\bf{H_1}}}({\bf{\widehat{{m}}}})||^p_p
 \end{equation} can be re-written as the weighted minimization \cite{Mohan2010}

 \begin{equation}
\min_{{\bf{\widehat{m}}}}  ||{\mathcal A}\left({\bf{\widehat{m}}}\right)-{\bf{\widehat{y}}}||^2_{\ell_2} + \lambda ~||{\bf{H_1}}({\bf\widehat{m}}) {\bf W}^{1/2}||^p_p
 \end{equation} where 
 \begin{equation}{\bf W}=[{{{\bf{H}}}_1^*(\bf\widehat{m})}{{{\bf{H}}}_1(\bf\widehat{m})}+~\epsilon~{\bf I}]^{{p/2}-1}.\end{equation} 
 
 Here, $\epsilon$ is a regularization parameter chosen for numerical stability and $\bf I$ is the identity matrix. Using the property $||{{{\bf{~H}}}_1(\bf\widehat{m})}||_F^2 \le ||{{{\bf{H}}}_1(\bf\widehat{m})} {\bf W}^{1/2}||_*,$ we can re-write the above nuclear norm minimization in terms of the Frobenius norm. Thus, the MUSSELS cost function in Eq. \ref{use Hankel} can be re-written using the IRLS formulation as ($C_3({\bf \hat m})$) \begin{equation}
\label{IRLS}
\widehat{\tilde{\bf~m}}=  \argmin_{{\bf{\widehat{m}}}}||{\mathcal{A}}\left({\bf{\widehat{m}}}\right)-{\bf{\widehat{y}}}||^2_{\ell_2}~+~\lambda~||{\bf{H_1}}({\bf{\widehat{{m}}}}){\bf W}^{1/2}||_F^2,
\end{equation}
where $_F$ represents the Frobenius norm. \\

\begin{algorithm}[h!]
\floatname{algorithm}{}
\renewcommand\thealgorithm{}
\begin{algorithmic}[1]
\STATE{ Initialize the algorithm by channel combining the measured k-space data to form ${\bf{\hat{m}}}^{(0)}= {\bf \cal{A}}^*{\bf \cal{A}}(\bf{\hat y})$. }
\STATE{ set $\lambda$}
\STATE{ set $n=1$}
\STATE{Repeat}
\STATE{~~~~~${\bf D}^{(n)}={\bf \bf{H_1}}({\bf \hat m}^{(n)})$ \text{ where } ${\bf{H_1}}$ \text{ is the block-Hankel matrix given in Eq. [\ref{equation 3}]}}.\\
\STATE{~~~~~Compute the singular value decomposition:  $\bf U\Sigma V^T=SVD( {\bf D}^*{\bf D})$}\\
\STATE{~~~~~Update ${\bf W}^{1/2}={\bf U\Sigma} ^{-1/4}$ (based on the property that $\bf {U\Sigma V}^T=\bf{US} ^2\bf{U}^T$)}\\
\STATE{~~~~~Update ${\bf{\hat{m}}}^{(n+1)}$ by miminizing  $C_3({\bf \hat m})$ given in Eq. [\ref{IRLS}] using CG }\\ 
~~~~~The gradient of $C_3({\bf \hat m})=2{\mathcal{A}}^*({\mathcal{A}}({\bf \hat m}^{(n)})-{\bf \hat y})+2{\lambda}{\bf H_1}^*\Big({ \bf H_1}({\bf \hat m}^{(n)}){{\bf W}^{1/2} {{\bf W}^{^*1/2}}}\Big)$\\
~~~~~Here ${\bf H_1}^{*}$ is the inverse mapping (adjoint ) of the block-Hankel elements into the multi-shot data matrix.
\STATE{ ~~~~~set $n=n+1$}
\STATE{ Until stopping criterion is reached}
\caption{Pseudo-code for IRLS MUSSELS}
\end{algorithmic}
\end{algorithm} 

 \vspace{-1.0em} \noindent { \it \large Advantages of IRLS formulation}\\ 

\vspace{-1.5em} \noindent The IRLS based MUSSELS formulation in Eq. \ref{IRLS} improves the performance of the reconstruction in several aspects. First of all, the new formulation eliminates the computationally expensive singular value shrinkage for rank minimization. Second, the new weight term $\bf W$ provides a denoiser. Thus, the new formulation alternates between a data consistency enforcement and a projection to the signal subspace, orthogonal to the null-space vectors specified by $\bf W$ during each iteration. The estimate of annihilation filters provided by $\bf W$ leads to faster recovery of the k-space samples. Note that the computation of $\bf W$ involves taking the inverse of a Gram matrix ${\bf{H}}_1^*(\bf\widehat{m}){{\bf{H}}}_1(\bf\widehat{m})$ which involves computing the SVD of ${\bf{H}}_1^*{\bf{H}}_1$. Since the size of ${\bf{H}}_1^*{\bf{H}}_1$ is $n \times n$, the computational complexity of the SVD is reduced to order $O(n^3)$. Assume that the SVD of ${\bf{H}}_1$ is given by $\bf USV^T$. Then, we evoke the property that SVD of ${\bf{H}}_1^*{\bf{H}}_1$ has the form $ \bf US^2U^T,$ to efficiently compute $\bf W$ (please see the pseudo-code). 
$\bf W$ is updated during the outer iteration and remains the same for the inner CG updates. Thus, we expect significant performance improvement in reconstructing the msDW data using the new IRLS formulation. In the present work, the estimate of the annihilation filters is performed from the data and is updated during every iteration. Nevertheless, the above formulation also allows the use of pre-estimated annihilation filters for faster recovery. A pseudo-code for minimizing the IRLS-based MUSSELS is given above.

\vspace{-0.0em} \noindent  Note that in this new framework, each column of  $\bf W$ is a null space vector or an annihilating filter. Because of the fact that the block-Hankel matrix is a concatenation of several Hankel matrices, the above formulation forms a multi-channel convolution where each column of $\bf W$ is a multi-channel filter. Figure 2 illustrates the multi-channel convolution underlying the above computation. The term ${{\bf H_1}^*\Big({ \bf H_1}({\bf \hat m}^{(n)}){{\bf W}^{1/2} {{\bf W}^{^*1/2}}}}\Big)$ performs a multi-channel convolution and deconvolution in a numerically efficient manner.\\

\vspace{-1.5em} \noindent We also note that the annihilating filter $\bf W$ is the k-space counterpart of the image-domain phase. Thus, without going back and forth from image-domain to frequency domain, the IRLS framework exploits the phase information in the frequency domain itself in a unified framework. \\

\vspace{-1.0em} \noindent { \it \large Extension of MUSSELS with conjugate symmetry constraints }\\ 
  
\vspace{-1.5em} \noindent  Diffusion MRI data are often collected with partial Fourier (pF) acceleration to keep the TE as low as possible. msDW data collected with partial Fourier acceleration are thus under-sampled by 55-65\%. The original MUSSELS implementation can recover pF-accelerated data via matrix completion involving several extended iterations. To reduce the blurring in the above reconstruction, the smoothness regularized MUSSELS (SR-MUSSELS) algorithm \cite{Mani2017a} was proposed to recover the under-sampled data exploiting sparsity like priors embedded in the structured low-rank matrix recovery. Alternatively, the conjugate symmetry (CS) property of the k-space data \cite{Blaimer2009} has been exploited by several authors \cite{Kim2016,Hu2018a} to recover pF data. This property can be easily accommodated into MUSSELS formulation also to reduce blurring and improve the reconstruction of pF data. We note that the annihilation relations (  Eq. \ref{sec_eq} ) that are valid on a given k-space data, are valid on its conjugate symmetric copy also. Hence, a new rank-deficient block-Hankel matrix ${\bf{H}}_1(\bf{\widehat{m}})$ of the form:  
\begin{equation}
\label{with CS}
{{{\bf{H}}}_1(\bf\widehat{m})}=\begin{bmatrix}{\bf{\cal{H}}}(\widehat{m_1})~~~{\bf{\cal{H}}}(\widehat{m_2})~~...~~{\bf{\cal{H}}}(\widehat{m_{N_s}})~~~{\bf{\cal{H}}}(\widehat{m_1}^\dagger)~~{\bf{\cal{H}}}(\widehat{m_2}^\dagger)~~...~~{\bf{\cal{H}}}(\widehat{m_{N_s}}^\dagger)\end{bmatrix}
\end{equation}
can exploit the CS property in MUSSELS reconstruction, where $\dagger$ represents the flipped conjugate of the k-space data. \\

 \vspace{-1.5em}\noindent Note that columns of ${\bf{H}}_1(\bf{\widehat{m}})$ are doubled when the CS property is exploited. The doubling of the shot-dimension will increase the computational complexity of original MUSSELS to $O(4mn^2)$ whereas it will only change the complexity of IRLS MUSSELS to $O(8n^3)$. While this will result in high computational demands in the original MUSSELS formulation, the burden on the IRLS formulation is minimal. Thus, the IRLS formulation provides a computationally efficient framework to work with higher shots and higher matrix sizes. It also offers the flexibility to incorporate additional constraints to improve the msDW reconstruction and potentially enable online reconstruction of the high-resolution DWIs from multi-shot acquisitions. In the following section, we test the proposed IRLS formulation on in-vivo data and compare it to state-of-the-art reconstruction methods.\\

\vspace{-1.0em} \noindent {\bf \Large Methods}\\ 
 
 \vspace{-1.5em}\noindent {\it \large Datasets}\\ 
 
\vspace{-1.5em} \noindent We employed a multi-shot EPI acquisition to collect high-resolution dMRI data. Two datasets were collected at two field strengths. The first dataset were acquired on the GE MR750W 3T scanner (maximum gradient amplitude of 33 mT/m and a maximum slew rate of 120 T/m/s), using a dual-spin echo diffusion sequence at 0.82$\times$0.82 mm in-plane resolution. Using 4-shot acquisition with pF at 59\%, the TE is 84ms for b-value of 700 s/$mm^2$. Other imaging parameters include FOV: 210$\times$210 mm, sampling matrix: 256$\times$152, slice thickness: 4mm and NEX=2. The second set of data were acquired on the GE MR950 7T scanner (maximum gradient amplitude of 50 mT/m and a maximum slew rate of 200 T/m/s). The data were collected using a 4-shot acquisition at 1.1 mm isotropic resolution with a b-value of 1000 s/$mm^2$. Since the T$_2^*$ decay is faster on 7T field strength, we employed a Stejskal-Tanner sequence to reduce the TE. Other imaging parameters include FOV: 210$\times$210 mm, sampling matrix: 192$\times$120,  slice thickness of 1.1 mm, pF at 62.5\% with TE = 53 ms. 100 slices spanning the whole brain were collected in an acquisition time of 40 mins. A comparison dataset at 2mm isotropic resolution was also collected using single-shot techniques. A dielectric pad was employed to improve the signal drop off towards the inferior brain regions. Both datasets consist of 60 diffusion directions and employed a 32-channel head coil. All experiments were performed on healthy volunteers following the Institutional Review Board requirements at the University of Iowa and obtaining informed written consent.\\

 \vspace{-1.5em}\noindent {\it \large Reconstruction of msDWI data}\\ 

\vspace{-1.5em} \noindent  To study the improvements offered by the IRLS MUSSELS implementation, the data were reconstructed using the original MUSSELS formulation without and with the conjugate symmetry lifting in Eq. \ref{with CS} and with the IRLS implementations without and with conjugate symmetry lifting. For ease of reference, the original MUSSELS implementation employing singular value shrinkage will be referred to as SVS MUSSELS and new formulation using the IRLS will be referred to as IRLS MUSSELS. A MUSE reconstruction was also implemented \cite{Chen2013a,Chu2015} to compare the computational performance of the reconstructions. All implementations were performed in Matlab (The MathWorks, Inc., Natick, MA, USA). 
The reconstruction was performed on a high-performance computing server having 48 cores with hyper-threading enabled and 256GB memory. After the reconstruction of DWIs, the datasets was co-registered for motion and eddy current correction using FSL's eddy\_correct (\url{https://fsl.fmrib.ox.ac.uk/fsl/fslwiki/FDT/UserGuide}). Following this, a single tensor fitting was performed to extract the primary diffusion directions for visualization of the improvement offered by the high-resolution images. The orientation distribution functions (ODF) were also computed from these datasets assuming a 3-fiber model, which helps to better visualize the improvements offered by the high-resolution images. \\

 \vspace{-1.5em}\noindent {\bf \large Results}\\ 
 
  \vspace{-1.5em} \noindent { \it \large Equivalence of SVS MUSSELS and IRLS MUSSELS}\\
  
   \vspace{-1.5em} \noindent  We first show the equivalence of the formulation in Eq. \ref{use Hankel} and \ref{IRLS}. As noted before, the MUSSELS reconstruction directly recovers the individual shot images which are combined using SOS to form the final magnitude DWI.  Figure  \ref{fig:fig3} shows the individual shot images and the magnitude DWI reconstructed using the SVS MUSSELS and IRLS MUSSELS from the first dataset. In spite of the fast reconstruction provided by the IRLS MUSSELS, the images are highly consistent across shots and the reconstruction methods, as expected. A color-coded fractional anisotropy (FA) map generated from all DWIs reconstructed using both the methods are also provided which demonstrates that the accelerated IRLS MUSSELS achieves the same reconstruction quality as the original MUSSELS implementation for all DWIs.\\

\vspace{-1.5em} \noindent { \it \large Improved partial Fourier reconstruction using conjugate symmetry property}\\
 
 \vspace{-1.5em} \noindent  
 
To study the improved recovery offered by incorporating the additional constraint of conjugate symmetry, we compare
 the reconstruction of the pF data without and with the constraint. Specifically, the former used the block-Hankel matrix given in Eq. \ref{equation 3}, while the latter employed the block-Hankel matrix given in Eq. \ref{with CS}.  Figure  \ref{fig:fig4} show the reconstructions from the SVS MUSSELS without and with CS constraint for a given diffusion weighting. For comparison, the MUSE reconstructions are also added. While the MUSE reconstruction is noisy (fig  \ref{fig:fig4}a), the SVS MUSSELS  provide robust reconstructions even without the CS constraint (fig  \ref{fig:fig4}b), thanks to its joint recovery of the k-space data \cite{Mani2017a}. The addition of the conjugate symmetry constraint further improved the SVS MUSSELS reconstruction (fig  \ref{fig:fig4}c) with the recovery of several fine anatomical details that were previously not visible in the reconstructions. Figure \ref{fig:fig5} show the FA maps that were computed using all the 60 DWIs for the above reconstruction methods, while figure S1 in the supporting information show the FA maps that were computed using the first 30 DWIs for the above reconstruction methods. It is observed that the SVS MUSSELS without CS is more robust to noise compared to MUSE and is further improved by the addition of the CS constraint. The IRLS MUSSELS also show similar trend. Figure  \ref{fig:fig6} show the color-coded FA maps derived from the 60 DWIs using the IRLS MUSSELS without and with CS. The zoomed color-coded FA maps also highlight the regions where the CS property provide more details in the images. For comparison, the FA maps from the MUSE reconstructions are also added in the figure.  \\

 \vspace{-1.5em}  \noindent Figure  \ref{fig:fig7} shows the reconstruction using various implementations of IRLS MUSSELS from dataset 2 that was collected on the 7T MRI.  Here, the partial Fourier acceleration is less severe compared to the first dataset; however, the dataset is noisier due to the smaller voxel volume. The IRLS MUSSELS reconstruction with CS provides sharper results for this dataset also with some improvement in the anatomical details. \\

\vspace{-1.5em} \noindent { \it \large Accelerated reconstruction}\\

  \vspace{-1.5em} \noindent Table 1 reports the time taken to reconstruct a given DWI using various implementations for both the datasets. For comparison, the MUSE reconstruction times for these datasets are reported in the last row. As noted from figures  \ref{fig:fig4}-\ref{fig:fig5}, the SVS MUSSELS without CS constraint provided robust reconstructions, however, the reconstruction time was higher than the MUSE Matlab implementation. From table 1, it can be seen that the IRLS formulation accelerated the above reconstruction by a factor of 3 (first and third row), making it faster than MUSE Matlab implementation for both the datasets tested. Similarly, the addition of CS constraint increased the reconstruction time of SVS MUSSELS by $\sim$ 4.5 times. The IRLS formulation accelerated the above reconstruction by about 6 times (second and fourth row). The reconstruction time for the IRLS MUSSELS with CS constraint is only modestly longer than MUSE reconstruction, but provides improved reconstruction. \\

\vspace{-1.5em} \noindent { \large \it Whole brain reconstruction}\\

\vspace{-1.5em} \noindent  The fast recovery enabled by the IRLS implementation makes it possible to reconstruct the whole brain high-resolution DWIs in a reasonable time. Since the data is highly parallelizable both along the diffusion directions and slices, we made use of parallel processing to reconstruct the whole brain dataset. Specifically, we parallelized the reconstruction of 6100 volumes (100 slices x 61 directions) using 48 hyper-threaded CPU cores that reconstructed the data in 1 hour. \\

 \vspace{-1.5em} \noindent   Figures \ref{fig:fig7}-\ref{fig:fig9} show the reconstruction of the 7T dataset using the proposed IRLS MUSSELS incorporating conjugate symmetry. Specifically, a single tensor fit and a q-ball based ODF assuming 3 fiber peaks were fitted to this data. A comparison of similar analysis performed on the 2mm isotropic data obtained on the same subject is also shown. In figure \ref{fig:fig8}, the primary diffusion direction recovered from the single tensor fit is overlaid on the apparent diffusion coefficient (ADC) map to better visualize the background contrast. Evidently, the high-resolution data follows the cortical folds and better represent the anisotropy in all brain voxels. In Figure  \ref{fig:fig9}, the plot of the ODFs from two comparable slices also show several remarkable differences between the two resolutions as highlighted.\\

\vspace{-1.0em} \noindent {\bf \Large Discussion}\\ 

 \vspace{-1.5em} \noindent As evident from figures 8-9, the high-resolution and low-resolution diffusion data are comparable in the regions that are primarily composed of homogeneous white matter voxels. For example, there is no notable difference within the major white matter bundles such as the corticospinal tract or the corpus callosum. However, the major differences in the single tensor representation and the multi-compartmental representations are in voxels composed of multiple tissues, i.e, in voxels closer to gray/white tissue boundaries, voxels with heterogeneous fiber orientations and those in the sub-cortical regions. The differences in such regions clearly point to the inadequacy of the low resolution to feature the subtle changes undergoing in those regions as a result of neuropathologies. Thus there is a clear biological motivation to push the spatial resolution of dMRI to exploit its superior sensitivity in detecting micro-structural changes \cite{Jones2018}. Here, we compare a 2mm isotropic data with a 1.1mm isotropic data; however, the recovery of even more anatomical details can be expected from sub-millimeter voxel resolutions in the range of .8 -.6 mm isotropic resolution.\\

 \vspace{-1.5em} \noindent The long echo-times associated with single-shot methods has been a barrier in achieving higher spatial resolutions in dMRI. The need for reducing the echo-time is extremely important in dMRI studies not only for achieving higher spatial resolution, but also for high b-value studies and has led to the development of dedicated hardware for diffusion studies \cite{Jones2018}. The high performing gradient sets utilized in the human connectome project \cite{Sotiropoulos2013}, the MGH-UCLA Skyra Connectome Scanner \cite{Setsompop2018a} and the head-only high gradient slew rate scanner at Mayo clinic \cite{Tan2016} are some examples. The improved gradient performance of these hardware provides the short TE and have pushed the spatial resolution to sub-millimeter levels already using single-shot techniques. However, to reduce the TE by half, the gradient performance requires double the gradient strength and a gradient rise time that is four times faster \cite{Heidemann2012}. Moreover, the physiological constraints impose hard limits on the maximum gradient amplitude and slew-rate that can be employed realistically. In contrast, such short TEs can be realized on clinical hardware using multi-shot techniques. Thus the synergistic combination of improved gradient hardware and multi-shot techniques can readily push the spatial resolution well below the standard 2mm isotropic resolution. \\

 \vspace{-1.5em} \noindent A major disadvantage of multi-shot EPI methods for diffusion imaging studies is related to the reconstruction of DWI that are not affected by motion-induced phase artifacts. Traditionally, such reconstructions were performed using navigator-based acquisition methods that provided an estimate of the motion-induced phase errors. Recently, navigator-free methods were also proposed to achieve this goal \cite{Chen2013a}. Here, we presented a fast and direct reconstruction of msDW data that is not affected by motion-induced phase artifacts. It is an improvement of the MUSSELS method that aims to reconstruct the individual shot images directly from the measured k-space data using parallel imaging. The immunity of MUSSELS reconstruction to motion-induced phase artifacts is similar to that of a SENSE reconstruction of individual shot-images. However, in contrast to the SENSE reconstruction of individual shot-images, the MUSSELS reconstruction  jointly recovers all the shot-images exploiting the low-rank prior on the block-Hankel matrix, ${\bf{H}}_1(\bf{\widehat{m}})$. \\
 
 \vspace{-1.5em} \noindent  In the original MUSSELS implementation, the reconstruction relied on singular value shrinkage for the low-rank recovery of the  structured block-Hankel matrix, ${\bf{H}}_1(\bf{\widehat{m}})$, to reconstruct the images. However, the high computational demands of the above method makes it inconvenient to be used for the reconstruction of whole brain high-resolution studies. In the present work, we re-formulate the MUSSELS recovery using an IRLS scheme that replaces the singular value shrinkage with a Frobenius norm such that fast recovery of the data is possible. The new formulation led to the acceleration of the MUSSELS reconstruction by several folds, improving the practical utility of the method for whole brain high-resolution studies. Here, we have demonstrated the feasibility of enabling routine high-resolution diffusion studies on clinical hardware using msDWI.  \\
 
 \vspace{-1.5em} \noindent  The acceleration offered by the IRLS method can be attributed to several factors. The new weighting term, $ \bf W$,  projects the block-Hankel matrix to a low-dimensional signal space. During each iteration, the estimate of $ \bf W$ is updated. From a computational perspective, the term $\bf W$ can be interpreted as a denoiser. The addition of this term differentiates the IRLS MUSSELS from the SVS MUSSELS conceptually, and improves the performance of the reconstruction. From an imaging physics perspective, $\bf W$ can also be interpreted as the representation of the motion-induced phase in the frequency domain. The implicit incorporation of the phase information accelerates the recovery problem. We also note that the incorporation of $ \bf W$ can be generalized to a broader setting. Whereas in the current setting, $ \bf W$ is learned from the data itself and updated during each iteration, it is possible to learn $ \bf W$ from exemplary data. Such approaches can enable learned formulation which can in-turn be exploited in a deep learning setting. Future methods can  enable such reconstructions of msDWI on the MRI scanners in a matter of milliseconds. \\

 \vspace{-1.5em} \noindent  Another disadvantage of multi-shot methods compared to single-shot methods is the increase in the scan time due to the increased number of TRs required to collect the multiple shots. One straight forward method to reduce the acquisition time is to perform in-plane acceleration either by skipping k-space lines in each shot or by skipping some shots altogether. The latter method has the advantage of reducing the number of TRs directly. The IRLS MUSSELS with CS proposed above can be used to recover such data. Another approach is to adopt slice acceleration using simultaneous multi-slice imaging (SMS). The utility of MUSSELS method to recover the slice aliased and phase corrupted msDW images were explored in a preliminary work \cite{Mani2019}. One main disadvantage of the above method is the long reconstruction time due to the increase in the data size from multiple slices. Future work will extend the IRLS MUSSELS method to support SMS acquisitions also. \\ 

\vspace{-1.5em} \noindent   The IRLS MUSSELS method presented here can be extended to several settings. For example, the IRLS MUSSELS with conjugate symmetry is conceptually similar to the SR-MUSSELS in terms of relying on an additional set of shot properties to recover the missing k-space data. While the SR-MUSSELS rely on the smoothness of the images to derive a set of annihilation relations on the derivative of the shot images, the conjugate symmetry property of the k-space data is exploited in the present work to derive an additional set of annihilation relations. The additional constraint is shown to improve the recovery of the partial Fourier data. This method can also enable the recovery of under-sampled multi-shot acquisitions. Similarly, the IRLS method can be extended to achieve faster implementation of SR-MUSSELS also. We also note that the image domain msDW recovery method proposed in \cite{Hu2018a} can also be efficiently implemented using the IRLS approach. \\

{\bf Conclusion}: We developed an improved MUSSELS reconstruction that can accommodate additional shot priors in a computationally efficient manner to recover DWIs with better accuracy from partial Fourier acquisitions. The proposed IRLS scheme reduces the computation time by a factor of 6 which significantly brings down the time involved in reconstructing whole brain high-resolution msDW data to enable routine studies.  \\

\vspace{2mm}
{\bf{{Acknowledgements}}}
\vspace{2mm}

Financial support for this study was provided by grants NIH 5 R01 EB022019, 5 R01 MH111578 and NIH 1R01EB019961-01A1. This work was conducted on MRI instruments funded by 1S10OD025025-01 and 1S10RR028821-01.

\newpage
 \begin{figure}[h]
\includegraphics[trim = 2mm 10mm 25mm 0mm, clip, width=1\textwidth]{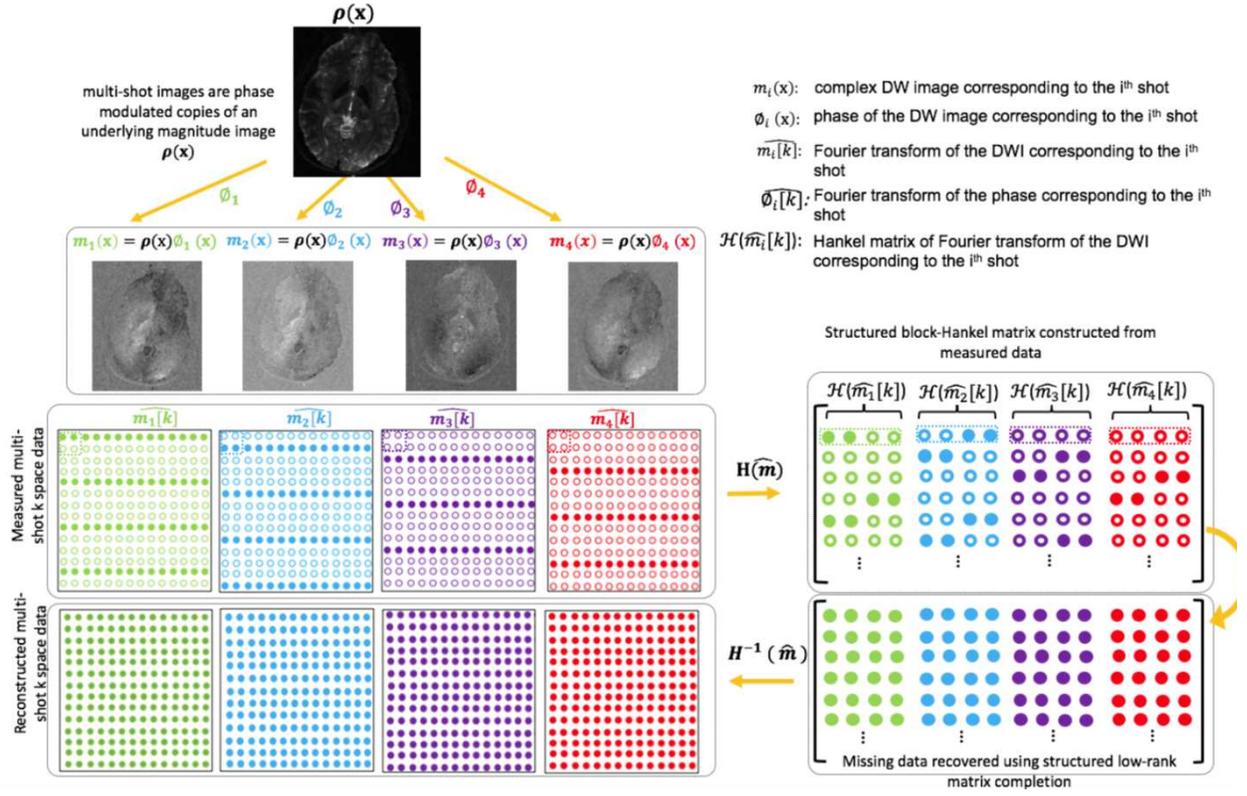}
\caption{Illustration of the direct reconstruction of msDW data using MUSSELS. $\rho(\bf{x})$ is the underlying magnitude DWI. The matrix of  multi-shot k-space data is represented as $\widehat{\bf{m[k]}}$. Solid and hollow circles denote respectively the measured and the missing k-space data in each shot. A block-Hankel matrix is created from the measured data whose missing samples are filled using matrix completion subject to data consistency. This, in turn, recovers the missing samples of the multi-shot k-space data . }
\label{fig:fig1}
\end{figure}
\clearpage

\newpage
 \begin{figure}[h]
\includegraphics[trim = 10mm 1mm 0mm 0mm, clip, width=1\textwidth]{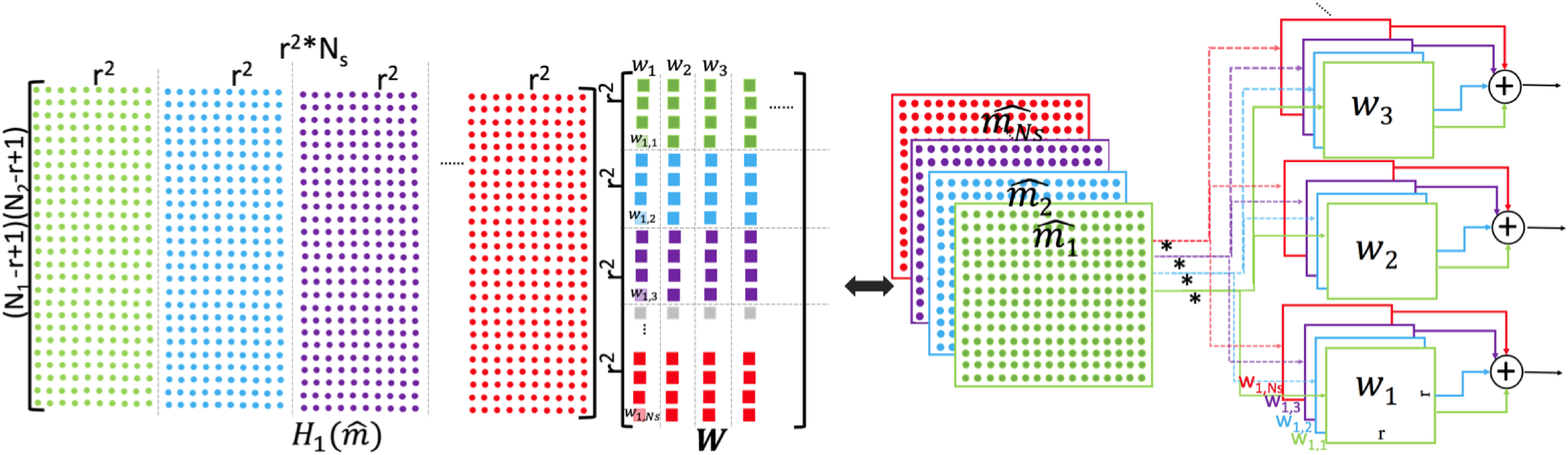}
\caption{Each column $ w_j$ of ${\bf W}$ is a multi-channel annihilation filter. The term ${\bf{H_1}}({\bf{\widehat{{m}}}}){\bf W}$ effectively computes a multi-channel convolution equivalent to the computation shown on the right side. Here, the k-space matrices $\hat{{\bf{m}}}$ are convolved with several multi-channel filters $w$ of size  $r \times r \times N_s$. }
\label{fig:fig2}
\end{figure}
\clearpage

\newpage
 \begin{figure}[h]
\includegraphics[trim = 10mm 20mm 0mm 0mm, clip, width=1\textwidth]{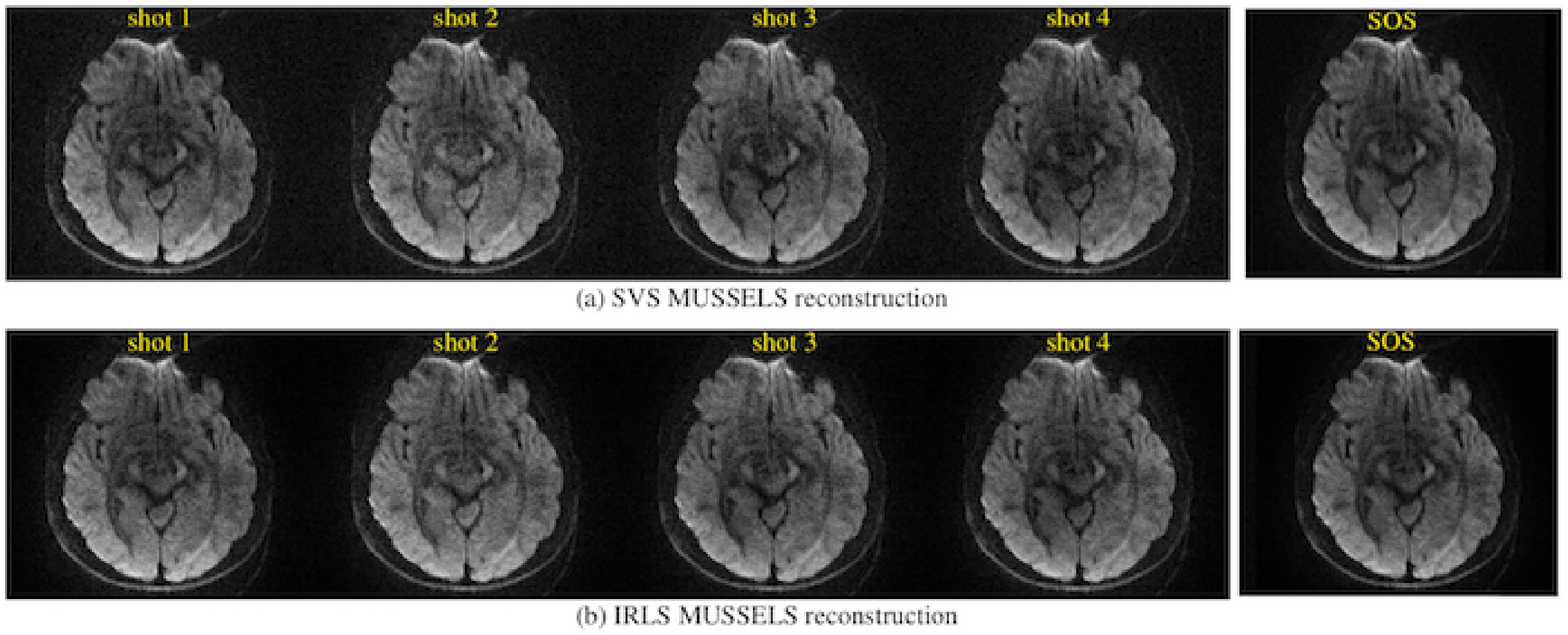}
\begin{minipage}[l]{0.6\textwidth}
\includegraphics[trim = 30mm 1mm 20mm 20mm, clip, width=1\textwidth]{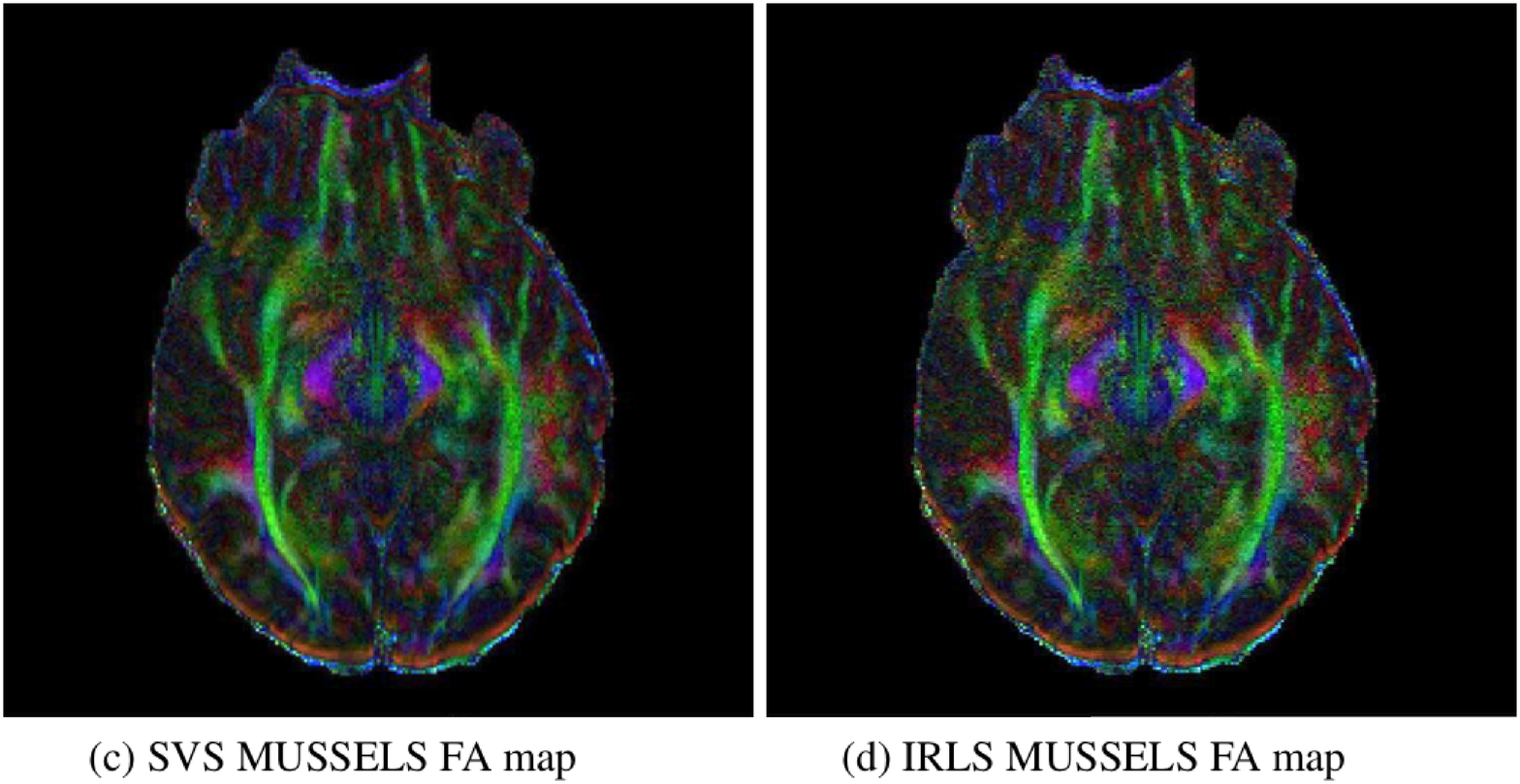}
\end{minipage}
\begin{minipage}[r]{0.4\textwidth}
\vspace{-4em}
\caption{Comparison of SVS MUSSELS reconstruction (a) and IRLS MUSSELS reconstruction (b). While formulation gives equivalent results, the IRLS MUSSELS reconstruction is much faster compared to the SVS MUSSELS reconstruction. (c) \& (d) shows the color coded directional FA map computed from all the DWIs corresponding to the two reconstructions. }
\label{fig:fig3}
\end{minipage}
\end{figure}
\clearpage


\newpage
 \begin{figure}[h]
\includegraphics[trim = 15mm 15mm 15mm 0mm, clip, width=1\textwidth]{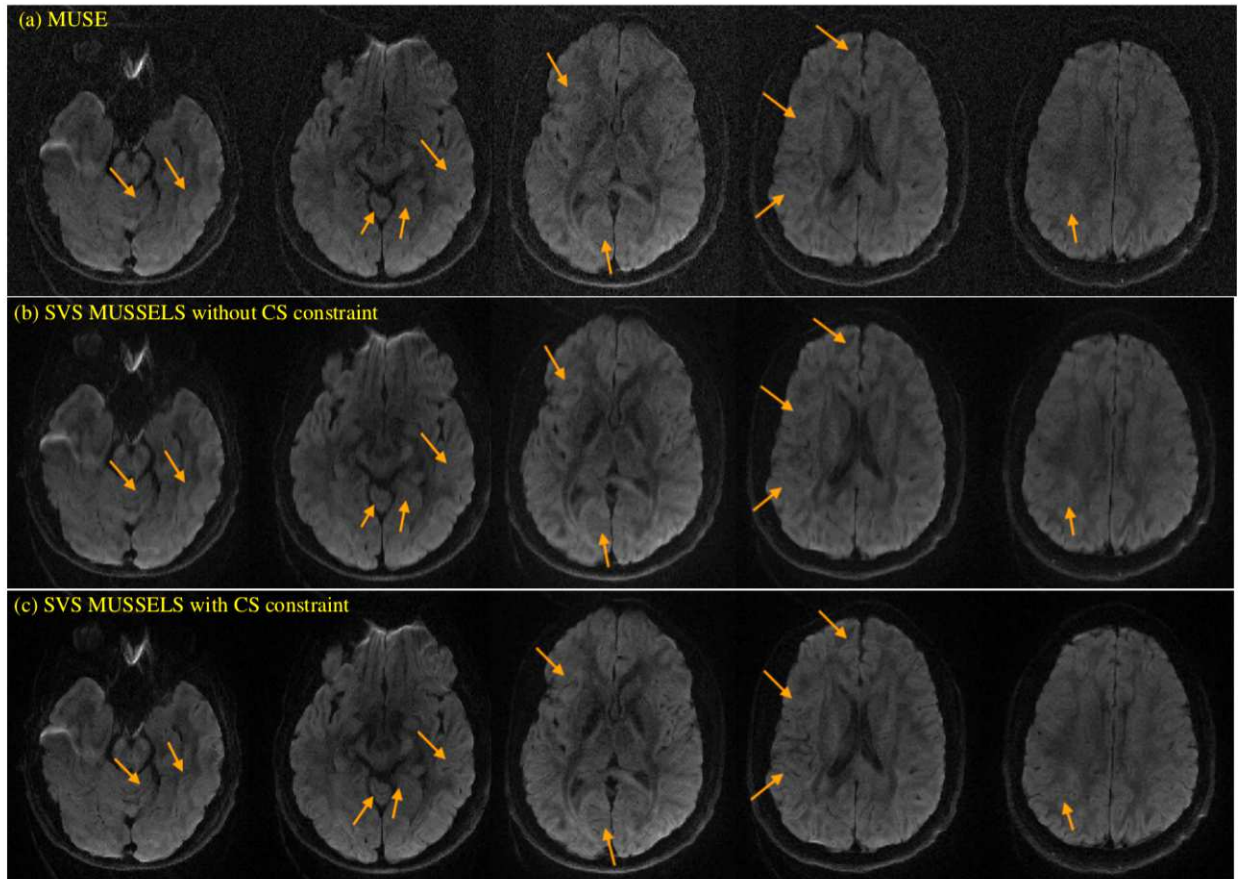}
\caption{DWI reconstruction from various slice locations show that the addition of conjugate symmetry constraint improves the recovery of anatomical details. The top row shows MUSE reconstruction, middle row shows SVS MUSSELS without CS, and the bottom row shows the SVS MUSSELS with CS constraint. Arrows highlight regions with better recovery of anatomical details.}
\label{fig:fig4}
\end{figure}
\clearpage


\newpage
 \begin{figure}[h]
\includegraphics[trim = 10mm 26mm 10mm 0mm, clip, width=1\textwidth]{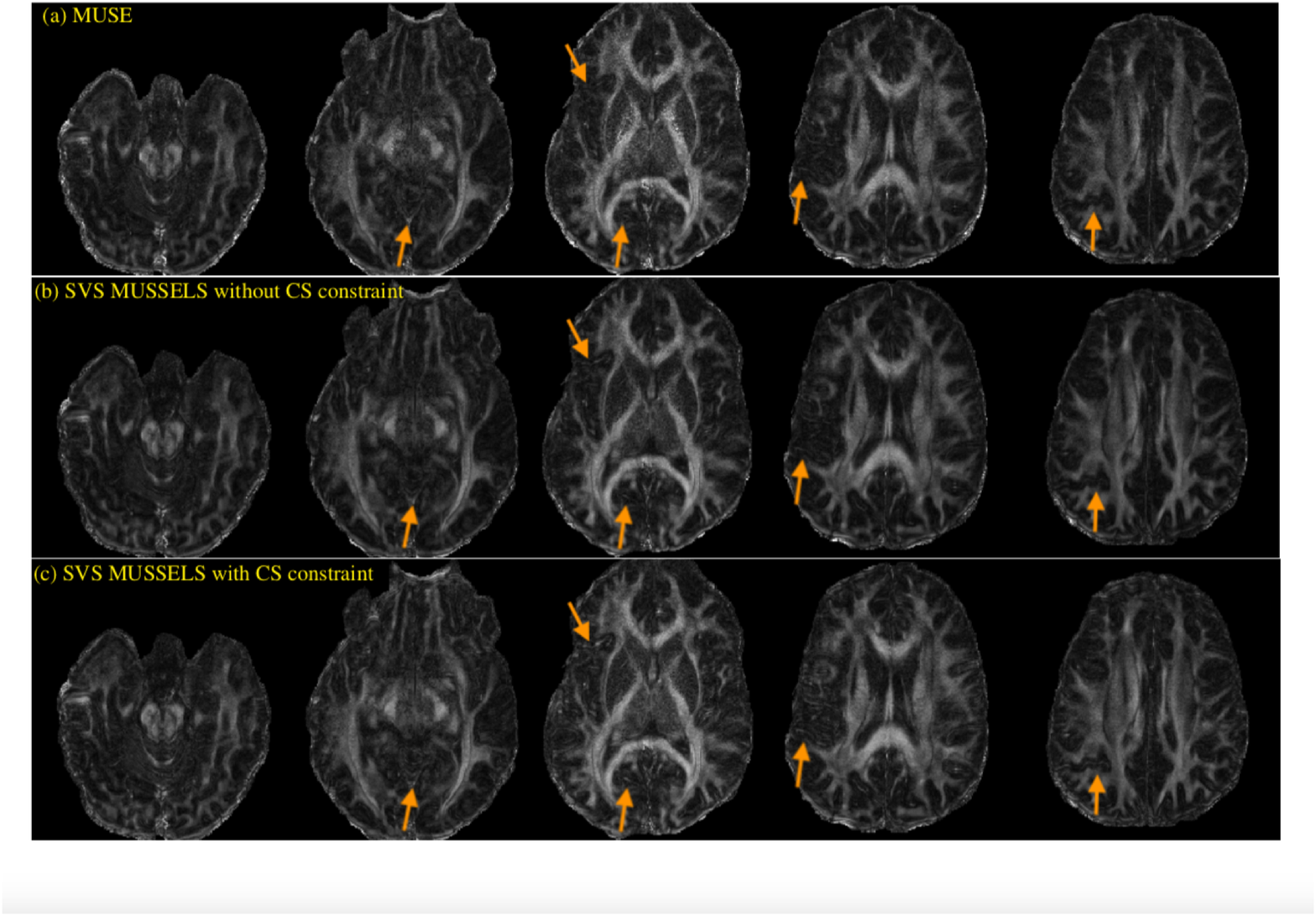}
\caption{FA maps reconstructed from several slice locations show that the addition of conjugate symmetry constraint improves the recovery of anatomical details (highlighted by arrows).   The top row shows the MUSE reconstruction, the middle row shows the SVS MUSSELS without CS, and the bottom row shows the SVS MUSSELS with CS constraint. }
\label{fig:fig5}
\end{figure}
\clearpage

\newpage
 \begin{figure}[h]
 \text{\rotatebox{90}{\fontsize{9}{5}\selectfont{ \hspace{0em}IRLS MUSSELS with CS \hspace{2em}IRLS MUSSELS without CS  \hspace{5em}MUSE}}}
\includegraphics[trim = 25mm 26mm 10mm 0mm, clip, width=1\textwidth]{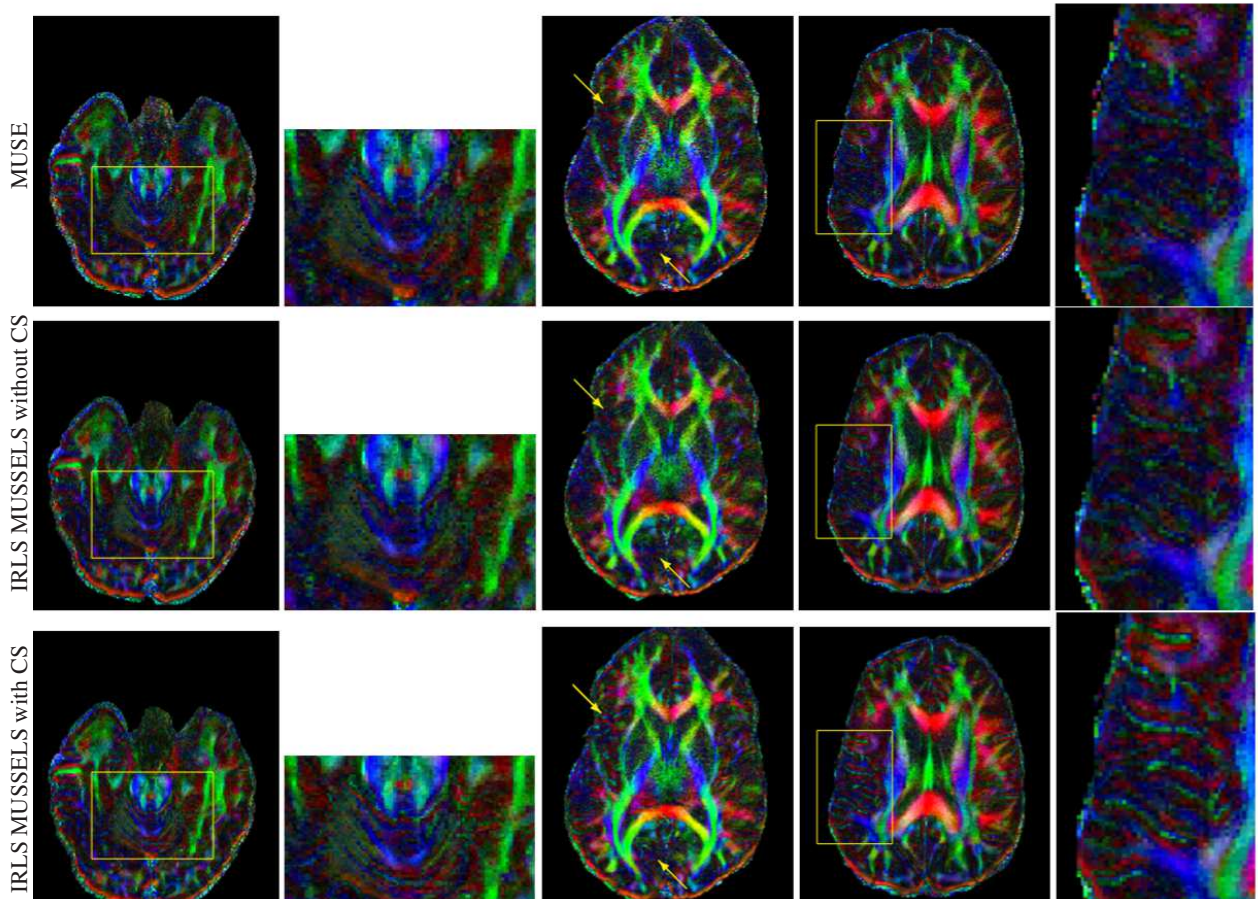}
\caption{Color-coded FA maps reconstructed from several slice locations for the various reconstructions. The top row shows the MUSE reconstruction, the middle row shows the IRLS MUSSELS without CS, and the bottom row shows the IRLS MUSSELS with CS constraint. Regions highlighted in yellow show that the addition of conjugate symmetry constraint improves the recovery of anatomical details Regions marked by the yellow boxes are zoomed to show the differences.   }
\label{fig:fig6}
\end{figure}
\clearpage

\newpage
 \begin{figure}[h!]
 \begin{minipage}[r]{0.65\textwidth}
\includegraphics[trim = 10mm 26mm 10mm 0mm, clip, width=1\textwidth]{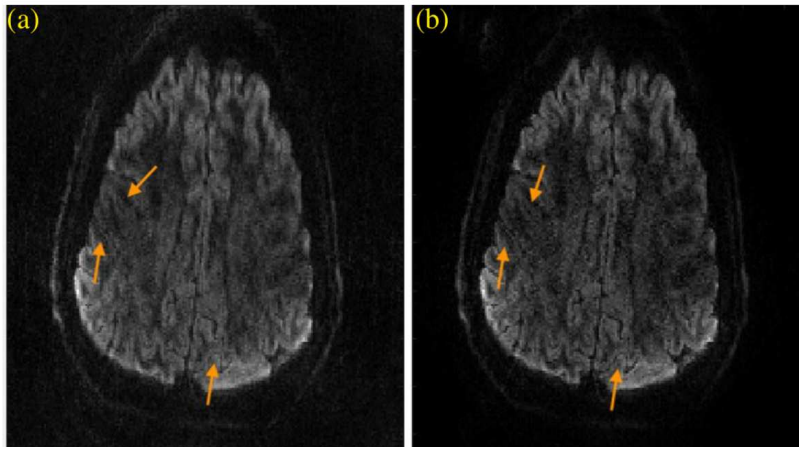}
\end{minipage}~
\begin{minipage}[r]{0.33\textwidth}
\vspace{-1em}
\caption{Comparison of IRLS MUSSELS reconstruction without (a) and with (b) CS performed on dataset 2. (b) shows sharper recovery of the data and the anatomical details are better defined than in (a) as indicated by the arrows. }
\label{fig:fig7}
\end{minipage}
\end{figure}
\clearpage

\newpage
 \begin{table}[ht]
\caption{Reconstruction time in seconds per image (per slice per direction)} 
\centering 
\begin{tabular}{c | c | c } 
\hline\hline 
Case &  3T dataset &7T dataset   \\ [0.5ex] 
~~~~ &   (256 $\times$ 256)  & (192 $\times$ 192)  \\ [0.5ex] 
\hline 

SVS MUSSELS &69 & 42\\ [1.5ex]
SVS MUSSELS with CS &322 & 190\\ [1.5ex]
IRLS MUSSELS &{\bf 24} & {\bf15} \\ [1.5ex]
IRLS MUSSELS with CS &50 &32 \\ [1.5ex]
MUSE & 40&25 \\ [1.5ex]
\end{tabular}
\label{table:nonlin} 
\end{table}

 \newpage
  \begin{figure}[h]
  \includegraphics[trim = 10mm 15mm 10mm 0mm, clip, width=1\textwidth]{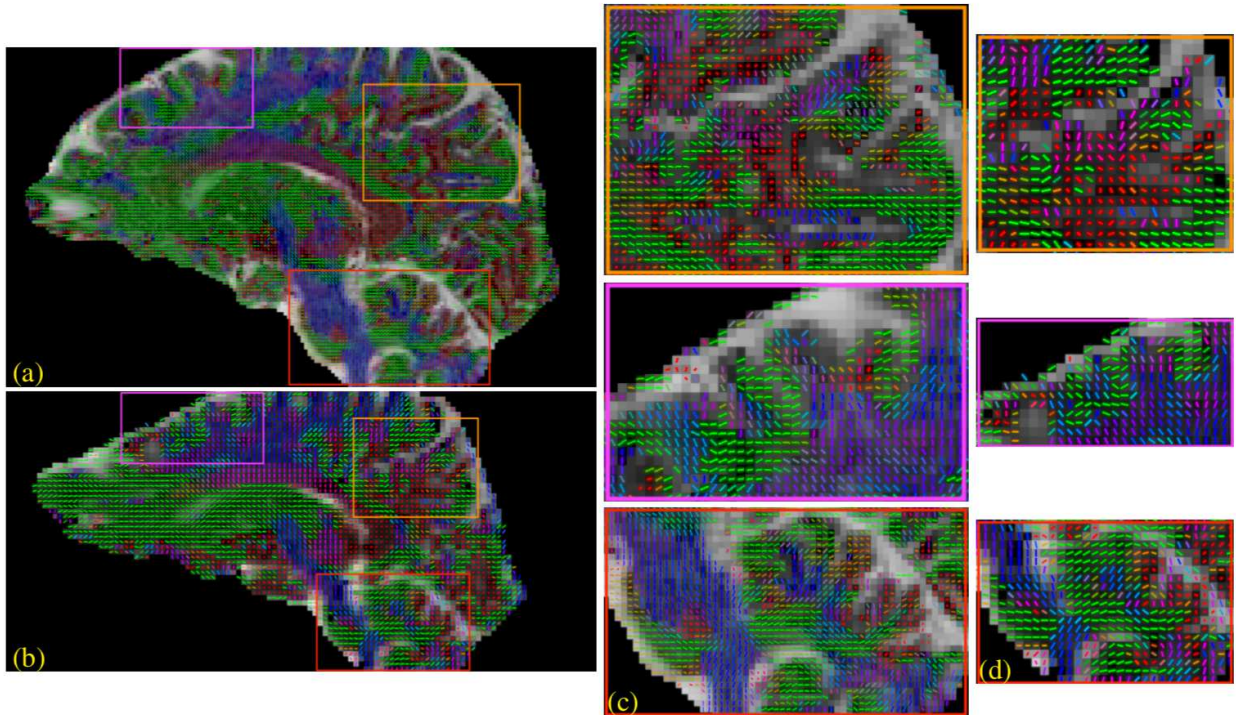}
\caption{Whole brain reconstruction of dataset 2. The IRLS with CS reconstruction was performed on the dataset and a single tensor model was fitted to the DWIs. This 1.1 mm isotropic dataset (a,c) is compared against the 2mm isotropic data (b,d) obtained from the same subject. Several regions are highlighted where the high-resolution data offers more details about the brain anatomy which is not fully captured by the low resolution data. (c) shows the zoomed view of regions highlighted in (a) and (d) shows the zoomed view of regions highlighted in (b).  }
\label{fig:fig8}
\end{figure}
\clearpage

 \newpage
 \begin{figure}[h!]
\includegraphics[trim = 5mm 10mm 5mm 0mm, clip, width=1\textwidth]{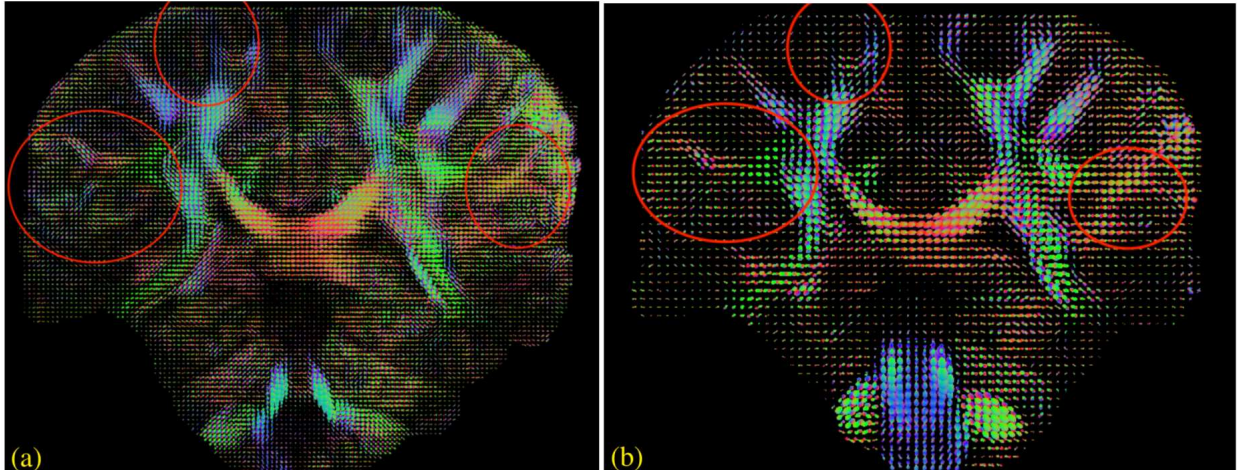}
\caption{Whole brain reconstruction from dataset 2. The ODFs reconstructed from the 1.1mm dataset (a) is compared to that of the 2mm dataset (b). Regions where the high-resolution data offers improved details about the brain anatomy compared to the low resolution data are highlighted. }
\label{fig:fig9}
\end{figure}
\clearpage

\newpage
{\bf{Legends: }}\\
Fig 1: Illustration of the direct reconstruction of msDW data using MUSSELS. $\rho(\bf{x})$ is the underlying magnitude DWI. The matrix of  multi-shot k-space data is represented as $\widehat{\bf{m[k]}}$. Solid and hollow circles denote respectively the measured and the missing k-space data in each shot. A block-Hankel matrix is created from the measured data whose missing samples are filled using matrix completion subject to data consistency. This, in turn, recovers the missing samples of the multi-shot k-space data . \\

Fig 2: Each column $ w_j$ of ${\bf W}$ is a multi-channel annihilation filter. The term ${\bf{H_1}}({\bf{\widehat{{m}}}}){\bf W}$ effectively computes a multi-channel convolution equivalent to the computation shown on the right side. Here, the k-space matrices $\hat{{\bf{m}}}$ are convolved with several multi-channel filters $w$ of size  $r \times r \times N_s$.\\

Fig 3: Comparison of SVS MUSSELS reconstruction (a) and IRLS MUSSELS reconstruction (b). While formulation gives equivalent results, the IRLS MUSSELS reconstruction is much faster compared to the SVS MUSSELS reconstruction. (c) \& (d) shows the color coded directional FA map computed from all the DWIs corresponding to the two reconstructions.\\

Fig 4: DWI reconstruction from various slice locations show that the addition of conjugate symmetry constraint improves the recovery of anatomical details. The top row shows MUSE reconstruction, middle row shows SVS MUSSELS without CS, and the bottom row shows the SVS MUSSELS with CS constraint. Arrows highlight regions with better recovery of anatomical details.\\

Fig 5: FA maps reconstructed from several slice locations show that the addition of conjugate symmetry constraint improves the recovery of anatomical details (highlighted by arrows).   The top row shows the MUSE reconstruction, the middle row shows the SVS MUSSELS without CS, and the bottom row shows the SVS MUSSELS with CS constraint.\\

Fig 6: Color-coded FA maps reconstructed from several slice locations for the various reconstructions. The top row shows the MUSE reconstruction, the middle row shows the IRLS MUSSELS without CS, and the bottom row shows the IRLS MUSSELS with CS constraint. Regions highlighted in yellow show that the addition of conjugate symmetry constraint improves the recovery of anatomical details Regions marked by the yellow boxes are zoomed to show the differences.\\

Fig 7: Comparison of IRLS MUSSELS reconstruction without (a) and with (b) CS performed on dataset 2. (b) shows sharper recovery of the data and the anatomical details are better defined than in (a) as indicated by the arrows.\\

Fig 8: Whole brain reconstruction of dataset 2. The IRLS with CS reconstruction was performed on the dataset and a single tensor model was fitted to the DWIs. This 1.1 mm isotropic dataset (a,c) is compared against the 2mm isotropic data (b,d) obtained from the same subject. Several regions are highlighted where the high-resolution data offers more details about the brain anatomy which is not fully captured by the low resolution data. (c) shows the zoomed view of regions highlighted in (a) and (d) shows the zoomed view of regions highlighted in (b). \\

Fig 9: Whole brain reconstruction from dataset 2. The ODFs reconstructed from the 1.1mm dataset (a) is compared to that of the 2mm dataset (b). Regions where the high-resolution data offers improved details about the brain anatomy compared to the low resolution data are highlighted.\\

Table 1: Reconstruction time in seconds per image (per slice per direction).\\

Fig S1: FA maps reconstructed from 60 DWIs (top row) and 30 DWIs (bottom row) using the various reconstruction methods. The reconstructed FA images from the 30 DWIs show that the MUSSELS reconstruction is more robust to the under-sampling of diffusion directions..\\

\newpage
{\small
\bibliographystyle{unsrtnat} 
\bibliography{IRLS_mussels}}

\begin{thebibliography}{30}
\expandafter\ifx\csname natexlab\endcsname\relax\def\natexlab#1{#1}\fi
\expandafter\ifx\csname url\endcsname\relax
  \def\url#1{{\tt #1}}\fi

\bibitem[Wu and Miller(2017)]{Wu2017}
Wu W and Miller KL.
\newblock {Image formation in diffusion MRI: A review of recent technical
  developments}.
\newblock {\em Journal of Magnetic Resonance Imaging}, 46\penalty0
  (3):\penalty0 646--662, 2017.

\bibitem[Novikov et~al.(2018)Novikov, Fieremans, Jespersen, and
  Kiselev]{Novikov2018b}
Novikov DS, Fieremans E, Jespersen SN, and Kiselev VG.
\newblock {Quantifying brain microstructure with diffusion MRI: Theory and
  parameter estimation}.
\newblock {\em NMR in Biomedicine}, page 3998, 2018.

\bibitem[Jones et~al.(2018)Jones, Alexander, Bowtell, Cercignani, Dell'Acqua,
  McHugh, Miller, Palombo, Parker, Rudrapatna, and Tax]{Jones2018}
Jones D, Alexander D, Bowtell R, Cercignani M, Dell'Acqua F, McHugh D, Miller
  K, Palombo M, Parker G, Rudrapatna U, and Tax C.
\newblock {Microstructural imaging of the human brain with a ‘super-scanner':
  10 key advantages of ultra-strong gradients for diffusion MRI}.
\newblock {\em NeuroImage}, 182:\penalty0 8--38, 2018.

\bibitem[Anderson and Gore(1994)]{Anderson1994}
Anderson AW and Gore JC.
\newblock {Analysis and correction of motion artifacts in diffusion weighted
  imaging.}
\newblock {\em Magnetic Resonance in Medicine}, 32\penalty0 (3):\penalty0
  379--87, 1994.

\bibitem[Eichner et~al.(2015)Eichner, Cauley, Cohen-Adad, M{\"{o}}ller, Turner,
  Setsompop, and Wald]{Eichner2015a}
Eichner C, Cauley SF, Cohen-Adad J, M{\"{o}}ller HE, Turner R, Setsompop K, and
  Wald LL.
\newblock {Real diffusion-weighted MRI enabling true signal averaging and
  increased diffusion contrast}.
\newblock {\em NeuroImage}, 122:\penalty0 373--384, 2015.

\bibitem[Butts et~al.(1996)Butts, de~Crespigny, Pauly, and Moseley]{Butts1996}
Butts K, de~Crespigny A, Pauly JM, and Moseley M.
\newblock {Diffusion-weighted interleaved echo-planar imaging with a pair of
  orthogonal navigator echoes.}
\newblock {\em Magnetic Resonance in Medicine}, 35\penalty0 (5):\penalty0
  763--70, 1996.

\bibitem[Chen et~al.(2013)Chen, Guidon, Chang, and Song]{Chen2013a}
Chen NK, Guidon A, Chang HC, and Song AW.
\newblock {A robust multi-shot scan strategy for high-resolution diffusion
  weighted MRI enabled by multiplexed sensitivity-encoding (MUSE)}.
\newblock {\em NeuroImage}, 2013.

\bibitem[Mani et~al.(2017)Mani, Jacob, Kelley, and Magnotta]{Mani2017a}
Mani M, Jacob M, Kelley D, and Magnotta V.
\newblock {Multi-shot sensitivity-encoded diffusion data recovery using
  structured low-rank matrix completion (MUSSELS)}.
\newblock {\em Magnetic Resonance in Medicine}, 78\penalty0 (2):\penalty0
  494--507, 2017.

\bibitem[Mani et~al.(2016)Mani, Magnotta, Kelley, and Jacob]{Mani2016d}
Mani M, Magnotta V, Kelley D, and Jacob M.
\newblock {Comprehensive reconstruction of multi-shot multi-channel diffusion
  data using mussels}.
\newblock In {\em Proceedings of the Annual International Conference of the
  IEEE Engineering in Medicine and Biology Society, EMBS}, 2016.
\newblock ISBN 9781457702204.

\bibitem[Hu et~al.(2018)Hu, Levine, Tian, Moran, Wang, Taviani, Vasanawala,
  McNab, Daniel, and Hargreaves]{Hu2018a}
Hu Y, Levine EG, Tian Q, Moran CJ, Wang X, Taviani V, Vasanawala SS, McNab JA,
  Daniel BA, and Hargreaves BL.
\newblock {Motion-robust reconstruction of multishot diffusion-weighted images
  without phase estimation through locally low-rank regularization}.
\newblock {\em Magnetic Resonance in Medicine}, 2018.

\bibitem[Holdsworth et~al.(2008)Holdsworth, Skare, Newbould, Guzmann, Blevins,
  and Bammer]{Holdsworth2008}
Holdsworth SJ, Skare S, Newbould RD, Guzmann R, Blevins NH, and Bammer R.
\newblock {Readout-segmented EPI for rapid high resolution diffusion imaging at
  3T}.
\newblock {\em European Journal of Radiology}, 2008.

\bibitem[Porter and Heidemann(2009)]{Porter2009}
Porter DA and Heidemann RM.
\newblock {High resolution diffusion-weighted imaging using readout-segmented
  echo-planar imaging, parallel imaging and a two-dimensional navigator-based
  reacquisition}.
\newblock {\em Magnetic Resonance in Medicine}, 62\penalty0 (2):\penalty0
  468--475, 2009.

\bibitem[Liu et~al.(2004)Liu, Bammer, Kim, and Moseley]{Liu2004}
Liu C, Bammer R, Kim Dh, and Moseley ME.
\newblock {Self-navigated interleaved spiral (SNAILS): Application to
  high-resolution diffusion tensor imaging}.
\newblock {\em Magnetic Resonance in Medicine}, 52\penalty0 (6):\penalty0
  1388--1396, 2004.

\bibitem[Guo et~al.(2016)Guo, Ma, Zhang, Zhang, Yuan, and Huang]{Guo2016}
Guo H, Ma X, Zhang Z, Zhang B, Yuan C, and Huang F.
\newblock {POCS-enhanced inherent correction of motion-induced phase errors
  (POCS-ICE) for high-resolution multishot diffusion MRI}.
\newblock {\em Magnetic Resonance in Medicine}, 75\penalty0 (1):\penalty0
  169--180, 2016.

\bibitem[Chu et~al.(2015)Chu, Chang, Chung, Truong, Bashir, and Chen]{Chu2015}
Chu ML, Chang HC, Chung HW, Truong TK, Bashir MR, and Chen NK.
\newblock {POCS-based reconstruction of multiplexed sensitivity encoded MRI
  (POCSMUSE): A general algorithm for reducing motion-related artifacts.}
\newblock {\em Magnetic Resonance in Medicine}, 74\penalty0 (5):\penalty0
  1336--48, 2015.

\bibitem[O'Halloran et~al.(2012)O'Halloran, Holdsworth, Aksoy, and
  Bammer]{OHalloran2012}
O'Halloran RL, Holdsworth S, Aksoy M, and Bammer R.
\newblock {Model for the correction of motion-induced phase errors in multishot
  diffusion-weighted-MRI of the head: Are cardiac-motion-induced phase errors
  reproducible from beat-to-beat?}
\newblock {\em Magnetic Resonance in Medicine}, 68\penalty0 (2):\penalty0
  430--440, 2012.

\bibitem[Pruessmann et~al.(1999)Pruessmann, Weiger, Scheidegger, and
  Boesiger]{Pruessmann1999}
Pruessmann KP, Weiger M, Scheidegger MB, and Boesiger P.
\newblock {SENSE: sensitivity encoding for fast MRI.}
\newblock {\em Magnetic resonance in medicine}, 42\penalty0 (5):\penalty0
  952--62, 1999.

\bibitem[Bertsekas(1976)]{Bertsekas1976}
Bertsekas DP.
\newblock {Multiplier methods: A survey}.
\newblock {\em Automatica}, 12\penalty0 (2):\penalty0 133--145, 1976.

\bibitem[Fornasier et~al.(2011)Fornasier, Rauhut, and Ward]{Fornasier2011}
Fornasier M, Rauhut H, and Ward R.
\newblock {Low-rank Matrix Recovery via Iteratively Reweighted Least Squares
  Minimization}.
\newblock {\em SIAM Journal on Optimization}, 21\penalty0 (4):\penalty0
  1614--1640, 2011.

\bibitem[Ramani and Fessler(2011)]{Ramani2011}
Ramani S and Fessler JA.
\newblock {Parallel MR image reconstruction using augmented Lagrangian
  methods.}
\newblock {\em IEEE Trans. Med. Imaging.}, 30\penalty0 (3):\penalty0 694--706,
  2011.

\bibitem[Ongie and Jacob(2017)]{Ongie2017}
Ongie G and Jacob M.
\newblock {A Fast Algorithm for Convolutional Structured Low-Rank Matrix
  Recovery}.
\newblock {\em IEEE Transactions on Computational Imaging}, 2017.

\bibitem[Chartrand and {Wotao Yin}(2008)]{Chartrand2008a}
Chartrand R and {Wotao Yin}.
\newblock {Iteratively reweighted algorithms for compressive sensing}.
\newblock In {\em 2008 IEEE International Conference on Acoustics, Speech and
  Signal Processing}, pages 3869--3872. IEEE, 2008.
\newblock ISBN 978-1-4244-1483-3.

\bibitem[Mohan and Fazel(2010)]{Mohan2010}
Mohan K and Fazel M.
\newblock {Iterative Reweighted Least Squares for Matrix Rank Minimization}.
\newblock In {\em 2010 48th Annual Allerton Conference on Communication,
  Control, and Computing (Allerton)}, 2010.
\newblock ISBN 9781424482160.

\bibitem[Blaimer et~al.(2009)Blaimer, Gutberlet, Kellman, Breuer,
  K{\"{o}}stler, and Griswold]{Blaimer2009}
Blaimer M, Gutberlet M, Kellman P, Breuer FA, K{\"{o}}stler H, and Griswold MA.
\newblock {Virtual coil concept for improved parallel MRI employing conjugate
  symmetric signals}.
\newblock {\em Magnetic Resonance in Medicine}, 61\penalty0 (1):\penalty0
  93--102, 2009.

\bibitem[Kim et~al.(2016)Kim, Setsompop, and Haldar]{Kim2016}
Kim TH, Setsompop K, and Haldar JP.
\newblock {LORAKS makes better SENSE: Phase-constrained partial fourier SENSE
  reconstruction without phase calibration.}
\newblock {\em Magnetic resonance in medicine}, 2016.

\bibitem[Sotiropoulos et~al.(2013)Sotiropoulos, Jbabdi, Xu, Andersson, Moeller,
  Auerbach, Glasser, Hernandez, Sapiro, Jenkinson, Feinberg, Yacoub, Lenglet,
  {Van Essen}, Ugurbil, and Behrens]{Sotiropoulos2013}
Sotiropoulos SN, Jbabdi S, Xu J, Andersson JL, Moeller S, Auerbach EJ, Glasser
  MF, Hernandez M, Sapiro G, Jenkinson M, Feinberg DA, Yacoub E, Lenglet C,
  {Van Essen} DC, Ugurbil K, and Behrens TEJ.
\newblock {Advances in diffusion MRI acquisition and processing in the Human
  Connectome Project}.
\newblock {\em NeuroImage}, 80:\penalty0 125--143, 2013.

\bibitem[Setsompop et~al.(2018)Setsompop, Fan, Stockmann, Bilgic, Huang,
  Cauley, Nummenmaa, Wang, Rathi, Witzel, and Wald]{Setsompop2018a}
Setsompop K, Fan Q, Stockmann J, Bilgic B, Huang S, Cauley SF, Nummenmaa A,
  Wang F, Rathi Y, Witzel T, and Wald LL.
\newblock {High-resolution in vivo diffusion imaging of the human brain with
  generalized slice dithered enhanced resolution: Simultaneous multislice
  (gSlider-SMS)}.
\newblock {\em Magnetic Resonance in Medicine}, 79\penalty0 (1):\penalty0
  141--151, 2018.

\bibitem[Tan et~al.(2016)Tan, Lee, Weavers, Graziani, Piel, Shu, Huston,
  Bernstein, and Foo]{Tan2016}
Tan ET, Lee SK, Weavers PT, Graziani D, Piel JE, Shu Y, Huston J, Bernstein MA,
  and Foo TK.
\newblock {High slew-rate head-only gradient for improving distortion in echo
  planar imaging: Preliminary experience}.
\newblock {\em Journal of Magnetic Resonance Imaging}, 44\penalty0
  (3):\penalty0 653--664, 2016.

\bibitem[Heidemann et~al.(2012)Heidemann, Anwander, Feiweier, Kn{\"{o}}sche,
  and Turner]{Heidemann2012}
Heidemann RM, Anwander A, Feiweier T, Kn{\"{o}}sche TR, and Turner R.
\newblock {k-space and q-space: combining ultra-high spatial and angular
  resolution in diffusion imaging using ZOOPPA at 7 T.}
\newblock {\em NeuroImage}, 60\penalty0 (2):\penalty0 967--78, 2012.

\bibitem[Mani et~al.(2019)Mani, Jacob, McKinnon, Yang, Rutt, Kerr, and
  Magnotta]{Mani2019}
Mani M, Jacob M, McKinnon G, Yang B, Rutt B, Kerr A, and Magnotta V.
\newblock {SMS MUSSELS: A Navigator-free Reconstruction for Simultaneous
  MultiSlice Accelerated MultiShot Diffusion Weighted Imaging}.
\newblock 2019.

\end{thebibliography}

\end{document}